\documentclass[prd,preprint,showpacs,superscriptaddress]{revtex4-1}

\usepackage{graphicx}
\usepackage{pdfpages}
\usepackage{hyperref}

\begin{document}
\title{VERITAS Deep Observations of the Dwarf Spheroidal Galaxy Segue 1}
\received{October 26$\mathrm{^{th}}$} \accepted{February $\mathrm{8^{th}}$}

\author{E.~Aliu}
\affiliation{Department of Physics and Astronomy, Barnard College, Columbia University, NY 10027, USA}
\author{S.~Archambault}
\affiliation{Physics Department, McGill University, Montreal, QC H3A 2T8, Canada}
\author{T.~Arlen}
\affiliation{Department of Physics and Astronomy, University of California, Los Angeles, CA 90095, USA}
\author{T.~Aune}
\affiliation{Santa Cruz Institute for Particle Physics and Department of Physics, University of California, Santa Cruz, CA 95064, USA}
\author{M.~Beilicke}
\affiliation{Department of Physics, Washington University, St. Louis, MO 63130, USA}
\author{W.~Benbow}
\affiliation{Fred Lawrence Whipple Observatory, Harvard-Smithsonian Center for Astrophysics, Amado, AZ 85645, USA}
\author{A.~Bouvier}
\affiliation{Santa Cruz Institute for Particle Physics and Department of Physics, University of California, Santa Cruz, CA 95064, USA}
\author{S.~M.~Bradbury}
\affiliation{School of Physics and Astronomy, University of Leeds, Leeds, LS2 9JT, UK}
\author{J.~H.~Buckley}
\affiliation{Department of Physics, Washington University, St. Louis, MO 63130, USA}
\author{V.~Bugaev}
\affiliation{Department of Physics, Washington University, St. Louis, MO 63130, USA}
\author{K.~Byrum}
\affiliation{Argonne National Laboratory, 9700 S. Cass Avenue, Argonne, IL 60439, USA}
\author{A.~Cannon}
\affiliation{School of Physics, University College Dublin, Belfield, Dublin 4, Ireland}
\author{A.~Cesarini}
\affiliation{School of Physics, National University of Ireland Galway, University Road, Galway, Ireland}
\author{J.~L.~Christiansen}
\affiliation{Physics Department, California Polytechnic State University, San Luis Obispo, CA 94307, USA}
\author{L.~Ciupik}
\affiliation{Astronomy Department, Adler Planetarium and Astronomy Museum, Chicago, IL 60605, USA}
\author{E.~Collins-Hughes}
\affiliation{School of Physics, University College Dublin, Belfield, Dublin 4, Ireland}
\author{M.~P.~Connolly}
\affiliation{School of Physics, National University of Ireland Galway, University Road, Galway, Ireland}
\author{W.~Cui}
\affiliation{Department of Physics, Purdue University, West Lafayette, IN 47907, USA }
\author{G.~Decerprit}
\affiliation{DESY, Platanenallee 6, 15738 Zeuthen, Germany}
\author{R.~Dickherber}
\affiliation{Department of Physics, Washington University, St. Louis, MO 63130, USA}
\author{J.~Dumm}
\affiliation{School of Physics and Astronomy, University of Minnesota, Minneapolis, MN 55455, USA}
\author{M.~Errando}
\affiliation{Department of Physics and Astronomy, Barnard College, Columbia University, NY 10027, USA}
\author{A.~Falcone}
\affiliation{Department of Astronomy and Astrophysics, 525 Davey Lab, Pennsylvania State University, University Park, PA 16802, USA}
\author{Q.~Feng}
\affiliation{Department of Physics, Purdue University, West Lafayette, IN 47907, USA }
\author{F.~Ferrer}
\affiliation{Department of Physics, Washington University, St. Louis, MO 63130, USA}
\author{J.~P.~Finley}
\affiliation{Department of Physics, Purdue University, West Lafayette, IN 47907, USA }
\author{G.~Finnegan}
\affiliation{Department of Physics and Astronomy, University of Utah, Salt Lake City, UT 84112, USA}
\author{L.~Fortson}
\affiliation{School of Physics and Astronomy, University of Minnesota, Minneapolis, MN 55455, USA}
\author{A.~Furniss}
\affiliation{Santa Cruz Institute for Particle Physics and Department of Physics, University of California, Santa Cruz, CA 95064, USA}
\author{N.~Galante}
\affiliation{Fred Lawrence Whipple Observatory, Harvard-Smithsonian Center for Astrophysics, Amado, AZ 85645, USA}
\author{D.~Gall}
\affiliation{Department of Physics and Astronomy, University of Iowa, Van Allen Hall, Iowa City, IA 52242, USA}
\author{S.~Godambe}
\affiliation{Department of Physics and Astronomy, University of Utah, Salt Lake City, UT 84112, USA}
\author{S.~Griffin}
\affiliation{Physics Department, McGill University, Montreal, QC H3A 2T8, Canada}
\author{J.~Grube}
\affiliation{Astronomy Department, Adler Planetarium and Astronomy Museum, Chicago, IL 60605, USA}
\author{G.~Gyuk}
\affiliation{Astronomy Department, Adler Planetarium and Astronomy Museum, Chicago, IL 60605, USA}
\author{D.~Hanna}
\affiliation{Physics Department, McGill University, Montreal, QC H3A 2T8, Canada}
\author{J.~Holder}
\affiliation{Department of Physics and Astronomy and the Bartol Research Institute, University of Delaware, Newark, DE 19716, USA}
\author{H.~Huan}
\affiliation{Enrico Fermi Institute, University of Chicago, Chicago, IL 60637, USA}
\author{G.~Hughes}
\affiliation{DESY, Platanenallee 6, 15738 Zeuthen, Germany}
\author{T.~B.~Humensky}
\affiliation{Physics Department, Columbia University, New York, NY 10027, USA}
\author{P.~Kaaret}
\affiliation{Department of Physics and Astronomy, University of Iowa, Van Allen Hall, Iowa City, IA 52242, USA}
\author{N.~Karlsson}
\affiliation{School of Physics and Astronomy, University of Minnesota, Minneapolis, MN 55455, USA}
\author{M.~Kertzman}
\affiliation{Department of Physics and Astronomy, DePauw University, Greencastle, IN 46135-0037, USA}
\author{Y.~Khassen}
\affiliation{School of Physics, University College Dublin, Belfield, Dublin 4, Ireland}
\author{D.~Kieda}
\affiliation{Department of Physics and Astronomy, University of Utah, Salt Lake City, UT 84112, USA}
\author{H.~Krawczynski}
\affiliation{Department of Physics, Washington University, St. Louis, MO 63130, USA}
\author{F.~Krennrich}
\affiliation{Department of Physics and Astronomy, Iowa State University, Ames, IA 50011, USA}
\author{K.~Lee}
\affiliation{Department of Physics, Washington University, St. Louis, MO 63130, USA}
\author{A.~S~Madhavan}
\affiliation{Department of Physics and Astronomy, Iowa State University, Ames, IA 50011, USA}
\author{G.~Maier}
\affiliation{DESY, Platanenallee 6, 15738 Zeuthen, Germany}
\author{P.~Majumdar}
\affiliation{Department of Physics and Astronomy, University of California, Los Angeles, CA 90095, USA}
\author{S.~McArthur}
\affiliation{Department of Physics, Washington University, St. Louis, MO 63130, USA}
\author{A.~McCann}
\affiliation{Physics Department, McGill University, Montreal, QC H3A 2T8, Canada}
\author{P.~Moriarty}
\affiliation{Department of Life and Physical Sciences, Galway-Mayo Institute of Technology, Dublin Road, Galway, Ireland}
\author{R.~Mukherjee}
\affiliation{Department of Physics and Astronomy, Barnard College, Columbia University, NY 10027, USA}
\author{R.~A.~Ong}
\affiliation{Department of Physics and Astronomy, University of California, Los Angeles, CA 90095, USA}
\author{M.~Orr}
\affiliation{Department of Physics and Astronomy, Iowa State University, Ames, IA 50011, USA}
\author{A.~N.~Otte}
\affiliation{Santa Cruz Institute for Particle Physics and Department of Physics, University of California, Santa Cruz, CA 95064, USA}
\author{N.~Park}
\affiliation{Enrico Fermi Institute, University of Chicago, Chicago, IL 60637, USA}
\author{J.~S.~Perkins}
\affiliation{CRESST and Astroparticle Physics Laboratory NASA/GSFC, Greenbelt, MD 20771, USA.}
\affiliation{University of Maryland, Baltimore County, 1000 Hilltop Circle, Baltimore, MD 21250, USA.}
\author{M.~Pohl}
\affiliation{Institut f\"ur Physik und Astronomie, Universit\"at Potsdam, 14476 Potsdam-Golm,Germany}
\affiliation{DESY, Platanenallee 6, 15738 Zeuthen, Germany}
\author{H.~Prokoph}
\affiliation{DESY, Platanenallee 6, 15738 Zeuthen, Germany}
\author{J.~Quinn}
\affiliation{School of Physics, University College Dublin, Belfield, Dublin 4, Ireland}
\author{K.~Ragan}
\affiliation{Physics Department, McGill University, Montreal, QC H3A 2T8, Canada}
\author{L.~C.~Reyes}
\affiliation{Physics Department, California Polytechnic State University, San Luis Obispo, CA 94307, USA}
\author{P.~T.~Reynolds}
\affiliation{Department of Applied Physics and Instrumentation, Cork Institute of Technology, Bishopstown, Cork, Ireland}
\author{E.~Roache}
\affiliation{Fred Lawrence Whipple Observatory, Harvard-Smithsonian Center for Astrophysics, Amado, AZ 85645, USA}
\author{H.~J.~Rose}
\affiliation{School of Physics and Astronomy, University of Leeds, Leeds, LS2 9JT, UK}
\author{J.~Ruppel}
\affiliation{Institut f\"ur Physik und Astronomie, Universit\"at Potsdam, 14476 Potsdam-Golm,Germany}
\affiliation{DESY, Platanenallee 6, 15738 Zeuthen, Germany}
\author{D.~B.~Saxon}
\affiliation{Department of Physics and Astronomy and the Bartol Research Institute, University of Delaware, Newark, DE 19716, USA}
\author{M.~Schroedter}
\affiliation{Fred Lawrence Whipple Observatory, Harvard-Smithsonian Center for Astrophysics, Amado, AZ 85645, USA}
\author{G.~H.~Sembroski}
\affiliation{Department of Physics, Purdue University, West Lafayette, IN 47907, USA }
\author{G.~D.~\c{S}ent\"{u}rk}
\affiliation{Physics Department, Columbia University, New York, NY 10027, USA}
\author{C.~Skole}
\affiliation{DESY, Platanenallee 6, 15738 Zeuthen, Germany}
\author{A.~W.~Smith}
\affiliation{Department of Physics and Astronomy, University of Utah, Salt Lake City, UT 84112, USA}
\author{D.~Staszak}
\affiliation{Physics Department, McGill University, Montreal, QC H3A 2T8, Canada}
\author{I.~Telezhinsky}
\affiliation{Institut f\"ur Physik und Astronomie, Universit\"at Potsdam, 14476 Potsdam-Golm,Germany}
\affiliation{DESY, Platanenallee 6, 15738 Zeuthen, Germany}
\author{G.~Te\v{s}i\'{c}}
\affiliation{Physics Department, McGill University, Montreal, QC H3A 2T8, Canada}
\author{M.~Theiling}
\affiliation{Department of Physics, Purdue University, West Lafayette, IN 47907, USA }
\author{S.~Thibadeau}
\affiliation{Department of Physics, Washington University, St. Louis, MO 63130, USA}
\author{K.~Tsurusaki}
\affiliation{Department of Physics and Astronomy, University of Iowa, Van Allen Hall, Iowa City, IA 52242, USA}
\author{A.~Varlotta}
\affiliation{Department of Physics, Purdue University, West Lafayette, IN 47907, USA }
\author{V.~V.~Vassiliev}
\affiliation{Department of Physics and Astronomy, University of California, Los Angeles, CA 90095, USA}
\author{S.~Vincent}
\affiliation{Department of Physics and Astronomy, University of Utah, Salt Lake City, UT 84112, USA}
\author{M.~Vivier}
\email{mvivier@bartol.udel.edu}
\affiliation{Department of Physics and Astronomy and the Bartol Research Institute, University of Delaware, Newark, DE 19716, USA}
\author{R.~G.~Wagner}
\affiliation{Argonne National Laboratory, 9700 S. Cass Avenue, Argonne, IL 60439, USA}
\author{S.~P.~Wakely}
\affiliation{Enrico Fermi Institute, University of Chicago, Chicago, IL 60637, USA}
\author{J.~E.~Ward}
\affiliation{School of Physics, University College Dublin, Belfield, Dublin 4, Ireland}
\author{T.~C.~Weekes}
\affiliation{Fred Lawrence Whipple Observatory, Harvard-Smithsonian Center for Astrophysics, Amado, AZ 85645, USA}
\author{A.~Weinstein}
\affiliation{Department of Physics and Astronomy, Iowa State University, Ames, IA 50011, USA}
\author{T.~Weisgarber}
\affiliation{Enrico Fermi Institute, University of Chicago, Chicago, IL 60637, USA}
\author{D.~A.~Williams}
\affiliation{Santa Cruz Institute for Particle Physics and Department of Physics, University of California, Santa Cruz, CA 95064, USA}
\author{B.~Zitzer}
\affiliation{Department of Physics, Purdue University, West Lafayette, IN 47907, USA}

\collaboration{The VERITAS collaboration}

\begin{abstract}
The VERITAS array of Cherenkov telescopes has carried out a deep observational program on the nearby dwarf spheroidal galaxy Segue 1. We report on the results of nearly 48 hours of good quality selected data, taken between January 2010 and May 2011. No significant $\gamma$-ray emission is detected at the nominal position of Segue 1, and upper limits on the integrated flux are derived. According to recent studies, Segue 1 is the most dark matter-dominated dwarf spheroidal galaxy currently known. We derive stringent bounds on various annihilating and decaying dark matter particle models. The upper limits on the velocity-weighted annihilation cross-section are $\mathrm{\langle\sigma v \rangle^{95\%\,CL} \lesssim 10^{-23}\,cm^{3}\,s^{-1}}$, improving our limits from previous observations of dwarf spheroidal galaxies by at least a factor of two for dark matter particle masses $\mathrm{m_{\chi}\gtrsim 300\,GeV}$. The lower limits on the decay lifetime are at the level of $\mathrm{\tau^{95\%\,CL} \gtrsim 10^{24}\,s}$.  Finally, we address the interpretation of the cosmic ray lepton anomalies measured by ATIC and PAMELA in terms of dark matter annihilation, and show that the VERITAS observations of Segue 1 disfavor such a scenario.\\
\end{abstract}


\maketitle
\section{Introduction}
The compelling evidence for the presence of non-baryonic dark matter  in various structures in the Universe \cite{DM} has motivated numerous efforts to search for dark matter.  Among many theoretical candidates for dark matter (see \cite{DM} for a review of candidates, and experimental searches), Weakly Interacting Massive Particles (WIMPs) are the most popular and well-motivated. A massive thermal relic of the early universe, with a weak scale interaction, naturally gives the measured present-day cold dark matter density $\mathrm{\Omega_{CDM}h^{2} = 0.1109 \pm 0.0056}$ \cite{RhoDM}. Candidates for WIMP dark matter are present in many extensions of the standard model of particle physics, such as supersymmetry (SUSY) \cite{SUSYDM} or theories with extra dimensions \cite{KKDM}. In such models, the WIMPs either decay or self-annihilate into standard model particles, which can produce a continuum of $\gamma$-rays with energies up to the dark matter particle mass, or monoenergetic $\gamma$-ray lines. Constraints from particle collider experiments are highly model-dependent, but generally place the mass of such particles in the range of a few tens of GeV to a few tens of TeV \cite{DM}. Indirect searches for dark matter with high energy (HE, $\mathrm{1\,GeV \leq E \leq 100\,GeV}$) or very high energy (VHE, $\mathrm{100\,GeV \leq E \leq 30\,TeV}$) $\gamma$-rays thus provide  a very promising way to test the nature of dark matter. Unlike cosmic ray charged particles, HE and VHE $\gamma$-rays are free of any propagation effects over short distances ($\mathrm{\leq}$ 1\,Mpc), and would therefore easily characterize a dark matter source location, spectrum and morphology. Such searches are generally conducted using pointed astrophysical observations of nearby dark matter overdensities, because the annihilation/decay rate strongly depends on the dark matter density. Popular targets include the Galactic Center \cite{GCBuckley, GCWhipple,GCHESS1,GCDM1, GCMAGIC, GCHESS2, GCDM2}, satellite galaxies of the Milky Way \cite{dSph1,dSph2,dSph3,dSph4,dSph5,dSphFermi,SegueFermi}, globular clusters \cite{GloC1,GloC2} and clusters of galaxy \cite{ClusterVERITAS,ClusterHESS,ClusterMAGIC, ClusterFermi1,ClusterFermi2,ClusterFermi3}. Non-targeted searches are also currently under consideration. They include blind searches for dark matter substructures in the galactic halo \cite{HESSIMBH,FERMIClumps,HESSClumps} and the measurement of the galactic and extra-galactic $\gamma$-ray diffuse emission \cite{ExtraGalacticDiffuse,FERMIAnisotropies1,FERMIDiffuse,FERMIAnisotropies2}. Compared to the targeted searches, non-targeted searches are much less affected by the uncertainties in the dark matter distribution modeling. However, they can suffer from large uncertainties in the modeling of the different backgrounds, especially in the HE $\gamma$-ray regime.\\
Beyond this well-established picture, additional effects might play an important role in the phenomenology of dark matter and the prospects for its detection. Motivated by the recent cosmic ray lepton spectra measured by the ATIC \cite{ATIC}, PAMELA \cite{PAMELApositronfraction,PAMELAelectron}, H.E.S.S. \cite{HESSelectron1,HESSelectron2} and Fermi-LAT \cite{Fermielectron1,Fermielectron2} experiments in the 1 GeV - 1 TeV energy range, various particle physics and astrophysics  effects have been suggested which could boost the dark matter signal with respect to standard expectations. For example, such particle physics effects include the Sommerfeld enhancement to the WIMP annihilation cross-section in the low dark matter particle velocity regime \cite{Sommerfeld1,Sommerfeld2,Sommerfeld3,AH}, or the internal bremsstrahlung effect \cite{IB1,IB2}, which can provide a considerable enhancement to the $\gamma$-ray signal at the endpoint of the spectrum. The presence of gravitationally-bound substructures within smooth dark matter halos can also have a significant impact on the annihilation/decay rate of dark matter particles \cite{Clumps1,Clumps2,Clumps3,Clumps4,Clumps5}. Such astrophysical enhancements, combined with particle physics enhancements, have been proposed as an explanation for the cosmic ray lepton anomalies \cite{Sommerfeld3,Clumps+PPE1,Clumps+PPE2}. Finally, decaying dark matter models have also been suggested to explain the lepton excesses and are good alternatives to annihilating dark matter models \cite{DDMPam1,DDMPam2,DDMPam3,DDMPam4,DDMPam5}.\\
The dwarf spheroidal galaxies (dSphs) of the Local Group best meet the criteria for a clear and unambiguous detection of dark matter. They are gravitationally-bound objects and are believed to contain up to $\cal{O}$($\mathrm{10^{3}}$) times more mass in dark matter than in visible matter, making them widely discussed as potential targets for indirect dark matter detection \cite{Clumps2,dSphDM1,dSphDM2,dSphDM3,dSphDM4,dSphDM5,dSphDM6}. Dwarf spheroidal galaxies are believed to be the remnants of dark matter halos, which contributed to the formation of Milky Way-sized galaxies during hierarchical clustering in structure formation scenarios. As opposed to the Galactic Center, and possibly globular clusters \cite{Terzan5,SgrDwarfProspect}, they are environments with a favorably low astrophysical $\gamma$-ray background. Neither astrophysical $\gamma$-ray sources (supernova remnants, pulsar wind nebulae, etc) nor gas acting as target material for cosmic rays have been observed in these systems \cite{dSphHI,dSphHI2}. Furthermore, their relative proximity and high galactic latitude make them the best astrophysical targets for high a signal-to-noise detection. With star velocity dispersions of the order of 10 $\mathrm{km\,s^{-1}}$ \cite{VelDisp}, dSphs are ideal laboratories for testing a possible velocity-dependence of the dark matter annihilation cross-section (for instance the Sommerfeld enhancement predicts $\mathrm{\langle\sigma v \rangle \propto 1/v}$). Like the Milky Way halo, dSphs are thought to harbor a population of substructures that could possibly boost their overall dark matter luminosity. However, recent simulations and analytic calculations show that for these objects, the expected astrophysical boost is less than a factor of ten \cite{Clumps3,Clumps4,dSphDM4,dSphDM5}.\\
The sensitivity improvement of the latest infrared/optical sky surveys has doubled the known number of Milky Way satellites in the past few years (for instance, see the Sloan Digital Sky Survey (SDSS) recent discoveries \cite{Segue1SDSS}), renewing the interest in these objects as promising targets for indirect dark matter searches. Segue 1 is one of these new Milky Way satellites, an ultra-faint dSph discovered in 2006 as an overdensity of resolved stars in the SDSS \cite{Segue1SDSS}. It is located at a distance of 23 $\mathrm{\pm}$ 2 kpc from the Sun at (RA,Dec) = (10$\mathrm{^{h}}$07$\mathrm{^{m}}$03.2$\mathrm{^{s}}$,16$\mathrm{^{\circ}}$04'25''), well above the galactic plane. Because of its proximity to the Sagittarius stream, the nature of the Segue 1 overdensity has recently been disputed, with some authors arguing that it was a tidally disrupted star cluster originally associated with the Sagittarius dSph \cite{Segue1GloC}. However, a kinematic study of a larger member-star sample (66 stars compared to the previous 24-star sample) has recently confirmed that Segue 1 is an ultra-faint Milky Way satellite galaxy \cite{Segue1dSph}. According to a study of its star kinematics, Segue 1 is probably one of the most dark matter-dominated dSph and is often highlighted as the most promising dSph target for indirect dark matter searches \cite{Segue1dSph,Segue1HL1,dSphDM4}. Segue 1 has been observed  in the HE $\gamma$-ray regime by the Fermi-LAT satellite \cite{FERMI} in its survey observation mode. Although no data analysis has been published yet by the Fermi-LAT collaboration, dedicated searches for dark matter using the first 9 months of public Fermi-LAT data on Segue 1 have been carried out by several authors \cite{SegueFermi,Segue1HL1}. No $\gamma$-ray signal was discovered in any of these analyses. The resulting upper limits on the dark matter velocity-weighted annihilation cross-section are at the level of $\mathrm{\langle\sigma v\rangle \sim 10^{-21}}$ - $\mathrm{10^{-23}\, cm^{3}\, s^{-1}}$ in the 10 GeV - 1 TeV WIMP mass range. In the VHE band, the MAGIC collaboration has recently conducted a search for dark matter annihilation in Segue 1, analyzing a 30-hour dataset taken in single telescope mode \cite{MAGICSegue1}. No VHE $\gamma$-ray signal was discovered, giving upper limits on $\mathrm{\langle\sigma v\rangle}$ at the level of $\mathrm{\sim 10^{-22}}$ - $\mathrm{10^{-23}\, cm^{3}\, s^{-1}}$ in the 100 GeV - 2 TeV WIMP mass range.\\
This paper reports on extensive observations of Segue 1 conducted by the Very Energetic Radiation Imaging Telescope Array System (VERITAS). After describing the VERITAS instrument, the observations and the data analysis in section \ref{ObsAna}, we extract integrated flux upper limits assuming various types of spectra in section \ref{ULs}. In section \ref{ClassicalDM}, bounds on annihilating and decaying dark matter are derived. Section \ref{leptons} addresses the interpretation of the recent cosmic ray lepton anomalies in terms of dark matter annihilation and presents the VERITAS $\gamma$-ray constraints using the Segue 1 data. Finally, section \ref{Conclusion} is devoted to our conclusions.
\section{Observations and analysis}\label{ObsAna}
\begin{figure*}[!ht]
   \centerline{\includegraphics[width=3in]{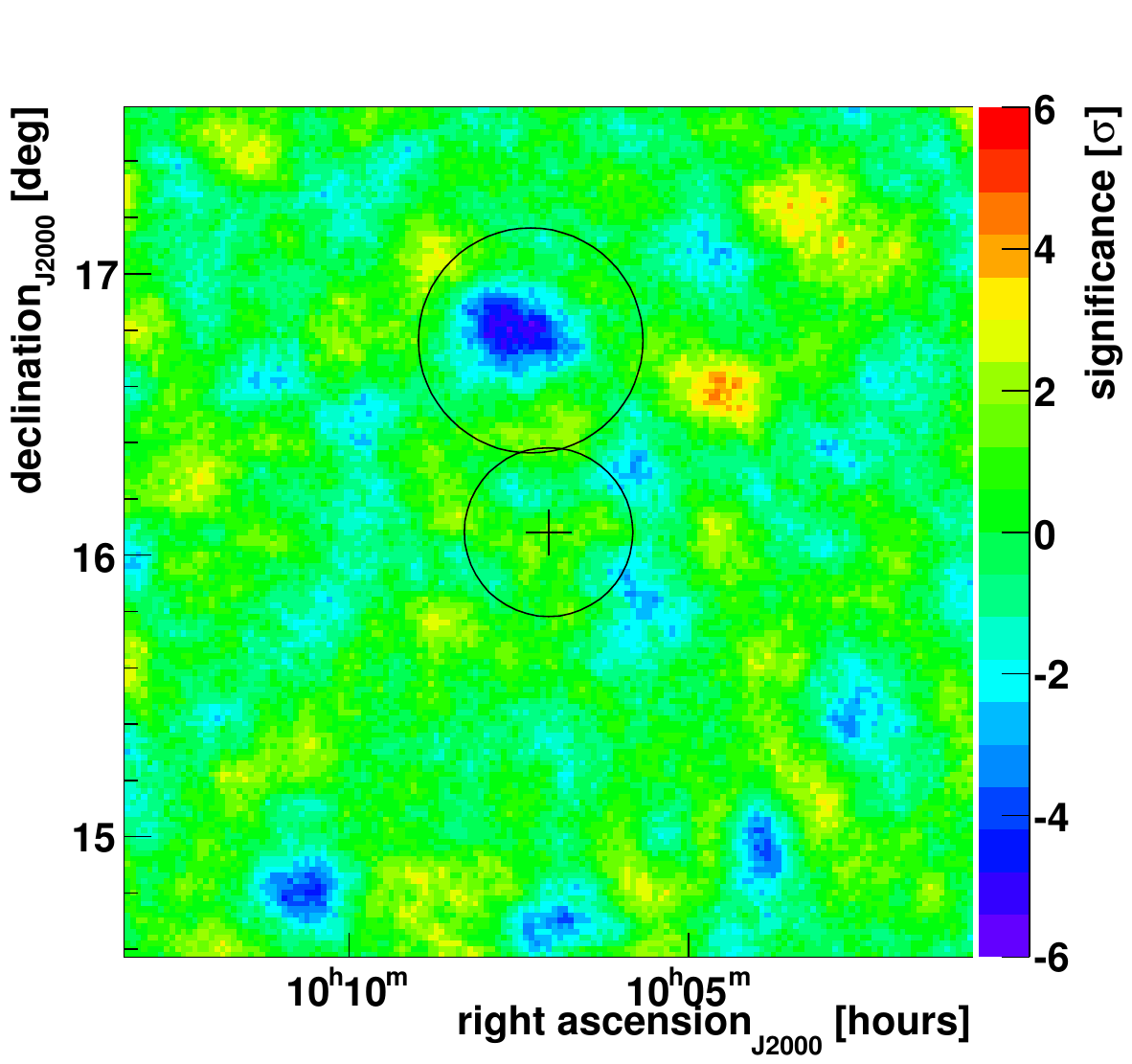}}
   \caption{Significance map obtained from the VERITAS observations of Segue 1 after $\gamma$-ray selection and background subtraction. The black cross indicates the position of Segue 1. The black circles correspond to the two exclusion regions used for the background determination. See text for further details.}
   \label{Fig1}
 \end{figure*}

VERITAS is an array of four 12-meter imaging atmospheric Cherenkov telescopes (IACTs) located at the base camp of the F. L. Whipple Observatory in southern Arizona (31$\mathrm{^{\circ}}$.68 N, 110$\mathrm{^{\circ}}$.95 W, 1.3 km above sea level). Each VERITAS telescope consists of a large optical reflector which focuses the Cherenkov light emitted by particle air showers onto a camera of 499 photomultiplier tubes. The array has been fully operational since September 2007. The large effective area ($\mathrm{\sim 10^{5}\,m^{2}}$), in conjunction with the stereoscopic imaging of air showers, enables VERITAS to be sensitive over a wide range of energies (from 100 GeV to 30 TeV) with an energy and angular resolution of 15\%-20\% and $\mathrm{0.1^{\circ}}$ per reconstructed $\gamma$-ray, respectively. VERITAS is able to detect a point source with 1\% of the Crab Nebula flux at a statistical significance of 5 standard deviations above background ($\mathrm{5\,\sigma}$) in approximately 25 hours of observations. For further details about VERITAS, see, e.g., \cite{VERITAS}.\\
Observations of the Segue 1 dSph were performed between January 2010 and May 2011. The data used for this analysis only includes observations taken under good weather and good data acquisition conditions. The total exposure of the dataset, after quality selection and dead time correction, amounts to 47.8 hours, and is the largest reported so far for any dSph observations conducted by an array of IACTs. The mean zenith angle of the observations is $\mathrm{\sim 20^{\circ}}$. The observations were conducted using the {\it wobble} pointing strategy, where the camera center is offset by $\mathrm{0.5^{\circ}}$ from the target position. The {\it wobble} mode allows for simultaneous background estimation and source observation, thus reducing the systematic uncertainties in the background determination \cite{Wobble}.\\
The data were reduced using standard VERITAS calibration and analysis tools \cite{VERITASdataanalysis}. After calibration of the photomultiplier tube gains \cite{VERITAScalibration}, images recorded by each of the VERITAS telescopes are characterized by a second-moment analysis giving the Hillas parameters \cite{HillasParameters1}. A stereoscopic analysis combining each telescope's image parameters is then used to reconstruct the $\gamma$-ray arrival direction and energy \cite{VERITASReconstruction}. We applied selection criteria (cuts) on the mean-reduced-scaled length and mean-reduced-scaled width parameters (see \cite{HillasParameters2} for a full description of these parameters) to reduce the hadronic cosmic ray background. The cuts for $\gamma$-ray/hadron separation were optimized {\it a priori} for a source with a 5\% Crab Nebula-like flux. Additionally, an event is accepted as a $\gamma$-ray candidate if the integrated charge recorded in at least two telescopes is $\mathrm{\geq}$ 90 photoelectrons, which effectively sets the analysis energy threshold to 170 GeV. Finally, a cut on $\mathrm{\theta}$, the angle between the target position and the reconstructed arrival direction, is applied to the $\gamma$-ray candidates and defines the signal search region ($\mathrm{\theta^{2}}$ $\mathrm{\leq}$ 0.015 deg$\mathrm{^{2}}$ in our analysis). After $\gamma$-ray selection, the residual background was estimated using the ring background technique \cite{RBck}. The ring background method computes the background for each position in the field of view using the background rate contained in a ring around that position. Two circular regions, of radius $\mathrm{0.2^{\circ}}$ centered on the target position and of radius $\mathrm{0.3^{\circ}}$ centered on the bright star $\eta$-Leonis (with apparent magnitude in the visible band $\mathrm{M_{V}}$ = 3.5, and located $\mathrm{0.68^{\circ}}$ from the position of Segue 1), were excluded for the background determination.\\
The analysis of the data resulted in the selection of $\mathrm{N_{ON}=1082}$ $\gamma$-ray candidates in the signal search region and $\mathrm{N_{OFF} = 12479}$ background events in the background ring region, with a normalization factor $\mathrm{\alpha=0.084}$, resulting in 30.4 excess events. The corresponding significance, calculated according to the method of Li \& Ma \cite{LiMa}, is $\mathrm{0.9\,\sigma}$. No significant $\gamma$-ray excess is found at the nominal position of Segue 1, nor in the whole field of view, as shown by the significance map on Figure \ref{Fig1}. The large depletion area, with negative significances, corresponds to the bright star $\eta$-Leonis.
\\
Given the absence of signal, one can derive upper limits (ULs) on the number of $\gamma$-rays in the source region. The computation of statistical ULs can be done following several methods, each relying on different assumptions. The bounded profile likelihood ratio statistic developed by Rolke et al. \cite{RolkeMethod} is used in our analysis. As discussed in the following sections, these ULs will serve for the computation of integrated flux ULs and for constraining some dark matter models. To make the computation of integrated flux ULs robust, we define a minimum energy, above which the energy reconstruction bias is less than 5\%. The energy reconstruction bias as a function of the reconstructed $\gamma$-ray energy has been studied with Monte Carlo simulations. The $\gamma$-ray selection cuts used in this analysis set the minimum reconstructed energy to $\mathrm{E_{min} = 300\, GeV}$. The ULs on the number of $\gamma$-rays computed with the Rolke prescription are displayed in Table \ref{Tab1}, along with the analysis results.
 \begin{table}[!ht]
\begin{center}
\begin{tabular}{|c|c|c|c|c|}
\hline
 Live time (min) & $\mathrm{E_{min}}$ (GeV) & $\mathrm{N^{exc}_{\gamma} (E\geq E_{min})}$ & Significance & $\mathrm{N_{\gamma}^{95\%\,CL} (E\geq E_{min})}$  \\
\hline
\hline
 2866 & - & 30.4 & 0.9 & 135.9\\
\hline
 2866 & 300 & 31.2 & 1.4 & 102.5\\
\hline
\end{tabular}
\end{center}
\caption{Analysis results of the VERITAS observations of Segue 1. $\mathrm{N^{exc}_{\gamma}(E\geq E_{min})}$ is the number of excess events in the signal search region with energies $\mathrm{E\geq E_{min}}$, after background subtraction. $\mathrm{N_{\gamma}^{95\%\,CL} (E\geq E_{min})}$ is the 95\% confidence level (CL) upper limit on the number of $\gamma$-rays with energies $\mathrm{E\geq E_{min}}$ in the signal search region, computed according to the Rolke \cite{RolkeMethod} prescription.}\label{Tab1}
\end{table}
\section{Flux upper limits}\label{ULs}
The analysis of the data did not show any significant excess over the background at the nominal position of Segue 1. The ULs on the number of $\gamma$-rays in the signal search region can then be converted to ULs on the integral $\gamma$-ray flux. The number of $\gamma$-rays detected by an array of IACTs above a minimum energy $\mathrm{E_{min}}$ is related to the source integral flux $\mathrm{\Phi_{\gamma}(E\geq E_{min})}$ by:
\begin{equation}
\mathrm{N_{\gamma}(E\geq E_{min})= T_{obs}\times\frac{\int_{E_{min}}^{\infty}{\cal{A}}_{eff}(E)\frac{dN_{\gamma}}{dE}dE}{\int_{E_{min}}^{\infty}\frac{dN_{\gamma}}{dE}dE}\times \Phi_{\gamma}(E\geq E_{min}),}\label{eqFluxUL}
\end{equation}
where $\mathrm{T_{obs}}$ is the observation time, $\mathrm{dN_{\gamma}/dE}$ the assumed source differential energy spectrum and $\mathrm{{\cal{A}}_{eff}(E)}$ is the instrument effective area. The effective area  $\mathrm{{\cal{A}}_{eff}(E)}$ is the instrument response function to the collection of $\gamma$-rays of energy E, and it depends on the zenith angle of the observations, the offset of the source from the target position and the $\gamma$-ray selection cuts. In the next two subsections, we consider two assumptions for the differential $\gamma$-ray spectrum: the case of a generic power-law spectrum, which describes well the TeV energy spectra of standard astrophysical sources and the case of a $\gamma$-ray spectrum resulting either from the annihilation or the decay of WIMP dark matter.
\subsection{Upper limits with power-law spectra}
Table \ref{Tab2} shows the integral flux ULs above $\mathrm{E_{min}=300\,GeV}$ for the assumption of a power-law spectrum:
\begin{equation}
\mathrm{\frac{dN_{\gamma}}{dE}\propto E^{-\Gamma},}
\end{equation} 
where $\mathrm{\Gamma}$ is the spectral index. The spectral indices have been varied over the range $\mathrm{\Gamma=1.8-3.0}$. The ULs on the integrated flux do not depend on the flux normalization (see eq. \ref{eqFluxUL}), but they do depend on the spectral index. A harder power-law spectrum provides a more constraining upper limit. The ULs on the integral flux reported in table \ref{Tab2} are at the level of 0.5\% of the Crab Nebula integral flux.
\begin{table}[!h]
\begin{center}
\begin{tabular}{|c|c|}
\hline
Spectral index & $\mathrm{\Phi_{\gamma}^{95\%\,CL}(E\geq 300\,GeV)}$ \\ 
$\mathrm{\Gamma}$ & [$\mathrm{10^{-13}\,cm^{-2}\,s^{-1}}$] \\
\hline
\hline
1.8 & 7.6 \\
2.2 & 7.7 \\
2.6 & 8.0 \\
3.0 & 8.2\\
\hline
\end{tabular}
\end{center}
\caption{95\% CL ULs on the integrated $\gamma$-ray flux above $\mathrm{E_{min} = 300\,GeV}$ from the VERITAS observations of  Segue 1, for power-law spectra with various spectral indices. For comparison, 1\% of the integrated Crab Nebula flux above $\mathrm{E_{min} = 300\,GeV}$ is $\mathrm{1.5\times 10^{-12}\,cm^{-2}\,s^{-1}}$.}\label{Tab2}
\end{table}
\subsection{Upper limits with Dark Matter $\gamma$-ray spectra}\label{FluxULsDM}
\begin{figure*}[!ht]
   \centerline{\includegraphics[width=3.3in]{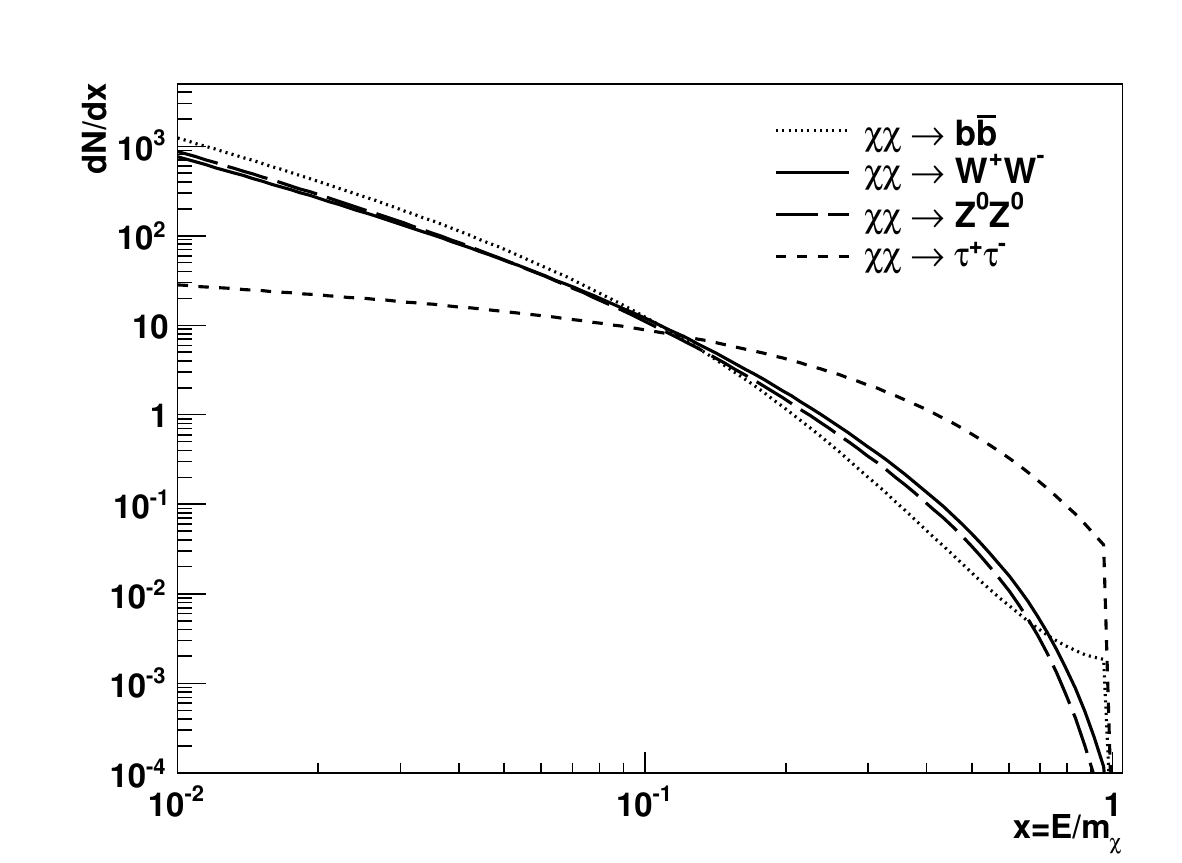}
              \hfil
              \includegraphics[width=3.3in]{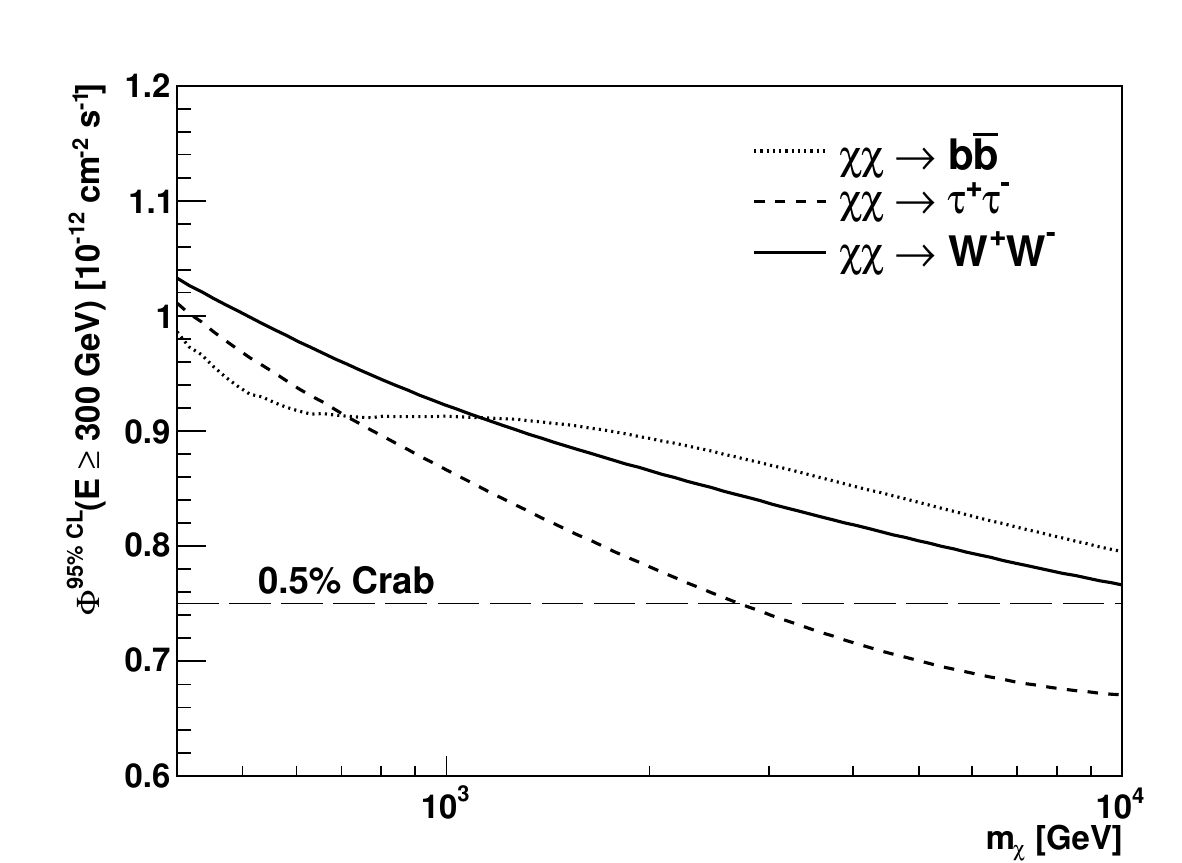}
             }
   \caption{Left: dark matter annihilation/decay spectra for four different final state products ($\mathrm{W^{+}W^{-}}$, $\mathrm{Z^{0}Z^{0}}$, $\mathrm{b\bar{b}}$ and $\mathrm{\tau^{+}\tau^{-}}$), extracted from PYTHIA 8.1 \cite{PYTHIA}. The spectra are plotted in the dN/dx representation, where $\mathrm{x=E/m_{\chi}}$ (or $\mathrm{x=2E/m_{\chi}}$ for decay spectra). Right: 95\% CL ULs on the integrated $\gamma$-ray flux above $\mathrm{E_{min} = 300\,GeV}$ from the VERITAS observations of Segue 1 considering dark matter particle annihilation/decay for three different channels:  $\mathrm{W^{+}W^{-}}$, $\mathrm{b\bar{b}}$ and $\mathrm{\tau^{+}\tau^{-}}$. For comparison, 0.5\% of the integrated Crab Nebula flux above $\mathrm{E_{min} = 300\,GeV}$ is $\mathrm{7.5\times 10^{-13}\,cm^{-2}\,s^{-1}}$.}
   \label{Fig2}
 \end{figure*}

Since Segue 1 is dark matter-dominated, one can derive ULs on the integrated flux assuming that the dominant source of $\gamma$-rays is dark matter annihilation or decay. The $\gamma$-ray differential energy spectrum from dark matter particle annihilation or decay depends on the dark matter model, and especially on the branching ratios to the final state particles. In almost every channel (excepting the $\mathrm{e^{+}e^{-}}$ and $\mathrm{\mu^{+}\mu^{-}}$ channels), the $\gamma$-ray emission mostly originates from the hadronization of the final state particles, with the subsequent production and decay of neutral pions. Annihilation/decay to three different final state products is considered independently of the dark matter model, in each case with a 100\% branching ratio: $\mathrm{W^{+}W^{-}}$, $\mathrm{b\bar{b}}$ and $\mathrm{\tau^{+}\tau^{-}}$. For each channel, the $\gamma$-ray spectrum has been simulated with the particle physics event generator PYTHIA 8.1 \cite{PYTHIA}. As shown by the left panel of Figure \ref{Fig2}, the $\mathrm{b\bar{b}}$ and $\mathrm{\tau^{+}\tau^{-}}$ channels encompass a wide range of dark matter $\gamma$-ray annihilation/decay spectra and give an idea of the uncertainties related to the dark matter particle physics model. The $\mathrm{Z^{0}Z^{0}}$ channel gives a $\gamma$-ray spectrum very similar to the $\mathrm{W^{+}W^{-}}$ channel and is not considered here. The right panel of Figure \ref{Fig2} shows the 95\% CL ULs on the integrated flux above $\mathrm{E_{min} = 300\,GeV}$ as a function of the dark matter particle mass. For $\mathrm{m_{\chi} \leq 800\,GeV}$, the most constraining ULs are obtained for the $\mathrm{b\bar{b}}$ channel because the $\mathrm{b\bar{b}}$ spectrum features a small ``bump'' at high $\mathrm{x=E/m_{\chi}}$ values (see Figure \ref{Fig2} left), which makes it the hardest among all the considered spectra. Increasing the dark matter particle mass, the x lower limit over which the $\gamma$-ray spectrum is integrated extends down to lower values, making the $\mathrm{b\bar{b}}$ $\gamma$-ray spectrum on average softer (and, inversely, the $\mathrm{\tau^{+}\tau^{-}}$ spectrum harder).  Above a dark matter particle mass of 800 GeV, the $\mathrm{\tau^{+}\tau^{-}}$ spectrum then gives the most constraining integrated flux ULs. The integrated flux ULs range between 0.3\% and 0.7\% of the Crab Nebula integral flux, depending on the dark matter particle mass.
\section{Dark Matter bounds}\label{ClassicalDM}
The absence of signal at the position of Segue 1 can be used to derive constraints on various dark matter models. Two different scenarios, in which the modeling of the $\gamma$-ray flux slightly differs, are considered: the case of annihilating dark matter and the case of decaying dark matter. 
\subsection{$\gamma$-rays from dark matter annihilation or decay}\label{DMModel}
The differential $\gamma$-ray flux from the annihilation of dark matter particles $\mathrm{\chi}$, of mass $\mathrm{m_{\chi}}$, in a spherical dark matter halo is given by a particle physics term multiplied by an astrophysical term \cite{dSphDM1}:
\begin{equation}
\mathrm{\frac{d\Phi_{\gamma}}{dE}(\Delta\Omega,E)=\frac{1}{4\pi}\frac{\langle\sigma v \rangle}{2\,m_{\chi}^{2}}\,\frac{dN_{\gamma}}{dE}\times\bar{J}(\Delta\Omega).}\label{dPhidE_DM}
\end{equation}
The particle physics term contains all the information about the dark matter model: the mass of the dark matter particle $\mathrm{m_{\chi}}$, the $\gamma$-ray differential energy spectrum from all final states weighted by their corresponding branching ratios $\mathrm{dN_{\gamma}/dE}$, and its total velocity-weighted annihilation cross-section $\mathrm{\langle\sigma v \rangle}$. The astrophysical factor $\mathrm{\bar{J}(\Delta\Omega)}$ (sometimes called the dark matter annihilation luminosity) is the square of the dark matter density integrated along the line of sight, s, and over the solid angle $\mathrm{\Delta\Omega}$ \cite{dSphDM1}:
\begin{equation}
\mathrm{\bar{J}(\Delta \Omega) = \int_{\Delta \Omega} d\Omega \int_{s_{min}}^{s_{max}}\rho_{\chi}^{2}(r[s])ds,}\label{fAP}
\end{equation}
where the upper and lower line of sight integration bounds depend on the distance d and the tidal radius $\mathrm{r_{t}}$ of the target:
\begin{equation}
\mathrm{s_{max,min} = d\,cos(\theta) \pm \sqrt{r_{t}^{2}-d^{2}\,\sin^{2}(\theta)}.}
\end{equation}
The solid angle $\mathrm{\Delta \Omega = 2\pi\,(1-cos(\theta_{max}))}$ is given here by the size of the signal search region defined previously in our analysis, i.e. $\mathrm{\theta^{2}_{max}}$ = 0.015 deg$\mathrm{^{2}}$. The estimate of the astrophysical factor requires a model of the Segue 1 dark matter profile. Motivated by results from N-body simulations, an Einasto profile \cite{Einasto1, Einasto2, Einasto3} is used:
\begin{equation}
\mathrm{\rho_{\chi}(r) = \rho_s\,e^{-2n\,[(r/r_s)^{1/n}-1]},} 
\end{equation}
with the scale density, the scale radius and the index n respectively being $\mathrm{\rho_s= 1.1\times 10^{8}\,M_{\odot}\,kpc^{-3}}$, $\mathrm{r_s = 0.15\,kpc}$ and n = 3.3 \cite{Private}. The value of the tidal radius changes the astrophysical factor by less than 10\%. We adopt a value of 500 pc, which is the median truncation radius of the Via Lactea II simulation subhalos presenting similar characteristics to the Segue 1 dark matter halo \cite{Segue1dSph,Clumps3}. Having these parameters in hand, the value of the astrophysical factor within the solid angle subtended by our on-source region is $\mathrm{\bar{J}(\Delta\Omega) = 7.7\times 10^{18}\,GeV^{-2}\,cm^{-5}\,sr}$. The systematic uncertainties on the astrophysical factor resulting from the fit of the Segue 1 dark matter distribution to an Einasto profile are less than an order of magnitude at the 1$\mathrm{\sigma}$ level \cite{Segue1HL1}.\\
In the scenario where the dark matter is a decaying particle, the expression of the differential $\gamma$-ray flux slightly differs and can be obtained with the following substitutions: 
\begin{itemize}
\item in eq. \ref{dPhidE_DM} $\mathrm{\langle\sigma v \rangle/2\,m_{\chi}^{2} \rightarrow \Gamma/m_{\chi}}$, where $\mathrm{\Gamma=\tau^{-1}}$ is the inverse of the dark matter particle decay lifetime.
\item in eq. \ref{fAP}, $\mathrm{\rho_{\chi}^{2} \rightarrow \rho_{\chi}}$.
\end{itemize}
Because particle decay is a one-body process, the astrophysical factor now depends on the dark matter density only. Adopting the same profile parameters, the value of the astrophysical factor in decaying dark matter scenarios is $\mathrm{\bar{J}(\Delta \Omega) = 2.6 \times 10^{17}\,GeV\,cm^{-2}\,sr}$. As in the annihilating dark matter case, the uncertainties on the astrophysical factor computed for an Einasto dark matter profile are less than one order of magnitude at the 1$\mathrm{\sigma}$ level \cite{MuonDecay2}.\\
The total number of $\gamma$-rays above a minimum energy $\mathrm{E_{\min}}$ expected in the signal source region is the integral of the differential $\gamma$-ray flux (eq. \ref{dPhidE_DM}), taking into account the instrument response to the collection of $\gamma$-rays:
\begin{equation}
\mathrm{N_{\gamma}(E\ge E_{min}) = T_{obs} \times \int_{E_{min}}^{\infty} {\cal{A}}_{eff}(E) \frac{d\Phi_{\gamma}}{dE}\,dE}\label{Ngamma}
\end{equation}
This equation is equivalent to eq. \ref{eqFluxUL} used for the calculation of integral flux upper limits (see section \ref{ULs}) if one assumes that the only source of $\gamma$-rays in Segue 1 is dark matter annihilation or decay. From eq. \ref{Ngamma} and taking $\mathrm{E_{min} = 0}$, the ULs on the number of $\gamma$-rays previously derived in section \ref{ObsAna} (see table \ref{Tab1}) can either be translated into ULs on the velocity-weighted annihilation cross-section $\mathrm{\langle\sigma v \rangle}$ or lower limits (LLs) on the decay lifetime $\mathrm{\tau}$.
\subsection{Upper limits on the annihilation cross-section}\label{sigmav}
\begin{figure*}[!ht]
   \centerline{\includegraphics[width=3.3in]{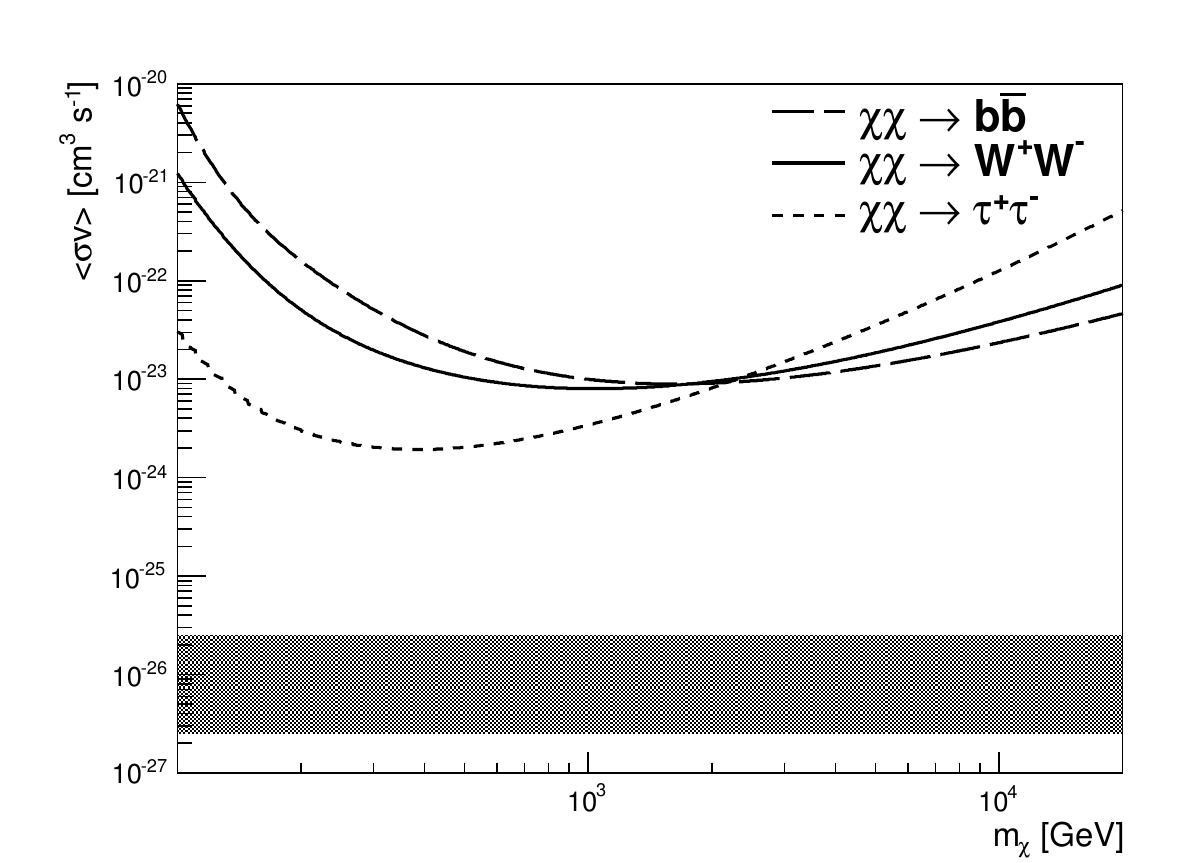}
              \hfil
              \includegraphics[width=3.3in]{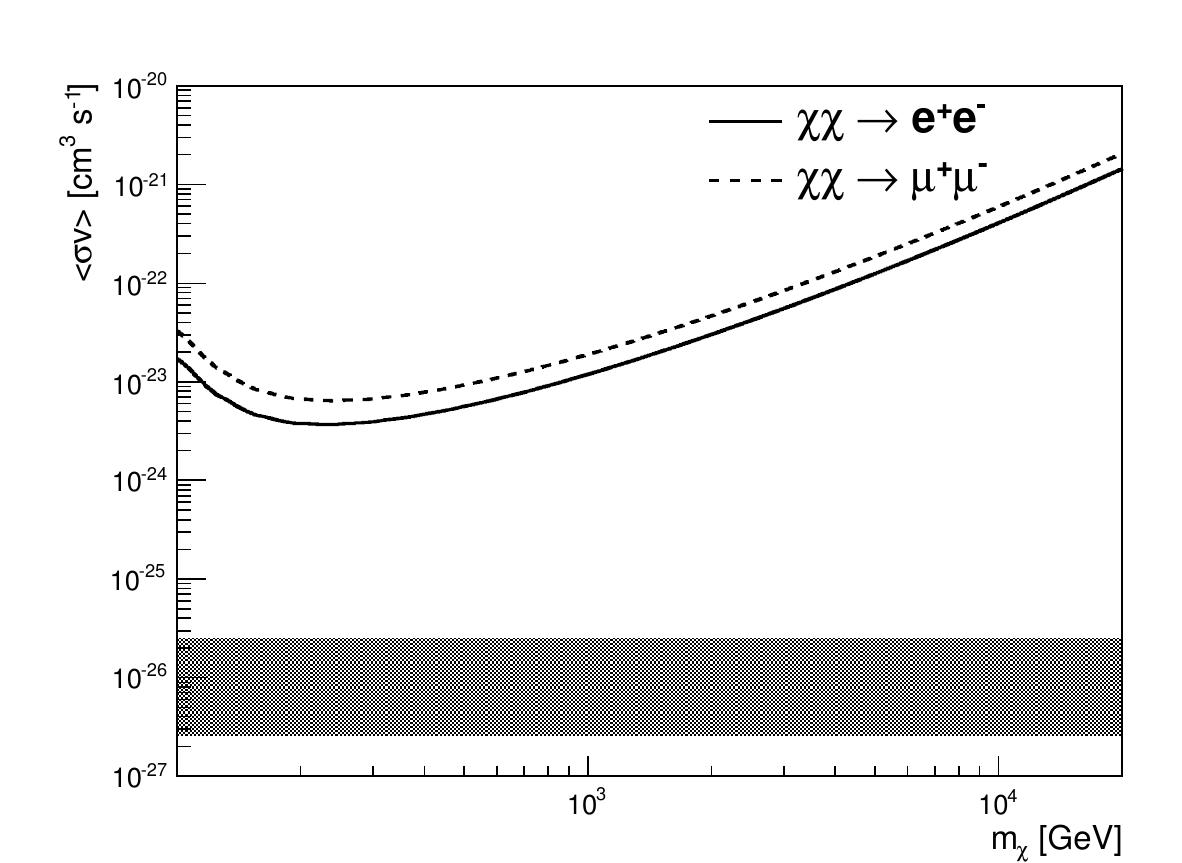}
             }
   \caption{95\% CL ULs from the VERITAS observations of Segue 1 on the WIMP velocity-weighted annihilation cross-section $\mathrm{\langle\sigma v \rangle}$ as a function of the WIMP mass, considering different final state particles. The grey band area represents a range of generic values for the annihilation cross-section in the case of thermally produced dark matter. Left: hadronic channels $\mathrm{W^{+}W^{-}}$, $\mathrm{b\bar{b}}$ and $\mathrm{\tau^{+}\tau^{-}}$. Right: leptonic channels $\mathrm{e^{+}e^{-}}$ and $\mathrm{\mu^{+}\mu^{-}}$.}
   \label{Fig3}
 \end{figure*}

In this section, ULs on the dark matter particle annihilation cross-section $\mathrm{\langle\sigma v \rangle}$ are derived independently of any dark matter models, considering five different final states with 100\% branching ratios: $\mathrm{W^{+}W^{-}}$, $\mathrm{b\bar{b}}$, $\mathrm{\tau^{+}\tau^{-}}$, $\mathrm{e^{+}e^{-}}$ and $\mathrm{\mu^{+}\mu{-}}$. The $\mathrm{W^{+}W^{-}}$, $\mathrm{b\bar{b}}$ and $\mathrm{\tau^{+}\tau^{-}}$ differential $\gamma$-ray spectra have been simulated with PYTHIA 8.1 \cite{PYTHIA} (see section \ref{FluxULsDM}) and are displayed on the left panel of Figure \ref{Fig2}. The final state radiation (FSR) $\mathrm{e^{+}e^{-}}$ and $\mathrm{\mu^{+}\mu{-}}$ channels are motivated by the recent anomalies measured in the cosmic ray lepton spectra (see section \ref{leptons}). The FSR spectrum for lepton channels has been computed analytically in \cite{IB1} for the $\mathrm{e^{+}e^{-}}$ case, and is given by:
\begin{equation}
\mathrm{\frac{dN}{dx} = \frac{\alpha}{\pi x}\biggl\{ \biggl[   (1-x)^{2} + 1 - \frac{m_{f}^{2}}{m_{\chi}^{2}}\biggr]\,ln\biggl(  \frac{1+\beta_f}{1-\beta_f} \biggr)  -2(1-x)\beta_f \biggr\},}\label{LeptonSpectra}
\end{equation}
where $\mathrm{\alpha}$ is the fine structure constant, $\mathrm{m_f}$ the fermion particle mass, $\mathrm{x=E/m_{\chi}}$ and $\mathrm{\beta_f = \sqrt{1-m_f^{2}/m_{\chi}^{2}\times(1-x)}}$. This analytical formula has been cross-checked against PYTHIA simulations and has given good agreement for both the $\mathrm{e^{+}e^{-}}$ and $\mathrm{\mu^{+}\mu^{-}}$ channels. In addition to the FSR contribution of the muons, the contribution of the radiative muon decay $\mathrm{\mu^{-} \rightarrow e^{-}\, \nu_{\mu}\, \bar{\nu}_{e}\, \gamma}$ and $\mathrm{\mu^{+} \rightarrow e^{+}\, \bar{\nu}_{\mu}\, \nu_{e}\, \gamma}$ has been included in the $\mathrm{\mu^{+}\mu^{-}}$ $\gamma$-ray spectra \cite{MuonDecay1,MuonDecay2}.\\
Figure \ref{Fig3} shows the 95\% CL exclusion curves on $\mathrm{\langle\sigma v \rangle}$ as a function of the dark matter particle mass for the five channels considered above, using eq. \ref{dPhidE_DM} and eq. \ref{Ngamma}. For the $\mathrm{W^{+}W^{-}}$ channel, the 95\% CL UL on the velocity-weighted annihilation cross-section is $\mathrm{\langle \sigma v \rangle^{95\%\,CL} \leq 8 \times 10^{-24}\, cm^{3}\, s^{-1}}$ at 1 TeV. This limit is the most constraining reported so far for any dSph observations in the VHE $\gamma$-ray band. The $\mathrm{b\bar{b}}$ and $\mathrm{\tau^{+}\tau^{-}}$ exclusion curves illustrate the range of uncertainties on the $\mathrm{\langle\sigma v \rangle}$ ULs from the dark matter particle physics model. Concerning the lepton channels $\mathrm{e^{+}e^{-}}$ and $\mathrm{\mu^{+}\mu^{-}}$, the limits are at the level of $\mathrm{10^{-23}\, cm^{3}\, s^{-1}}$ at 1 TeV. The current ULs on $\mathrm{\langle \sigma v \rangle}$ are two orders of magnitude above the predictions for thermally produced WIMP dark matter.
\subsection{Lower limits on the decay lifetime}
\begin{figure*}[!ht]
   \centerline{\includegraphics[width=3.3in]{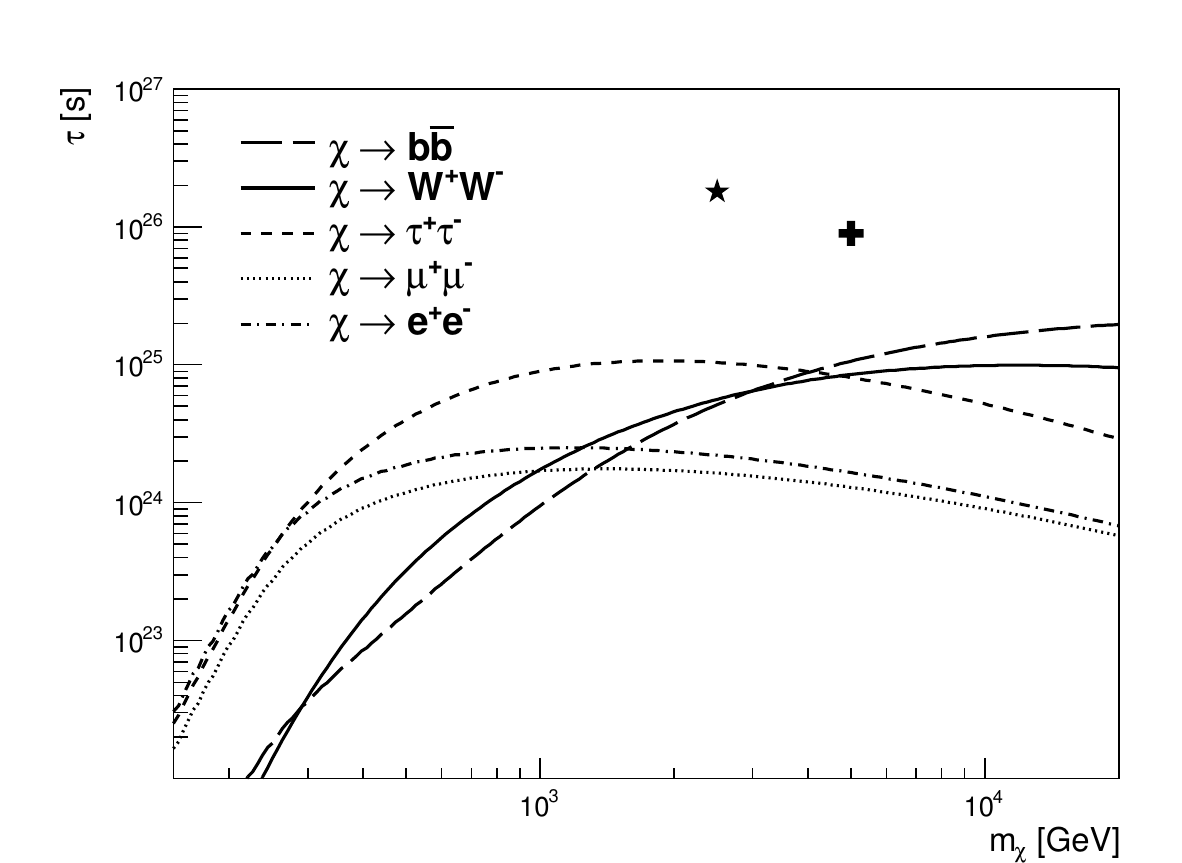}
              \hfil
              \includegraphics[width=3.3in]{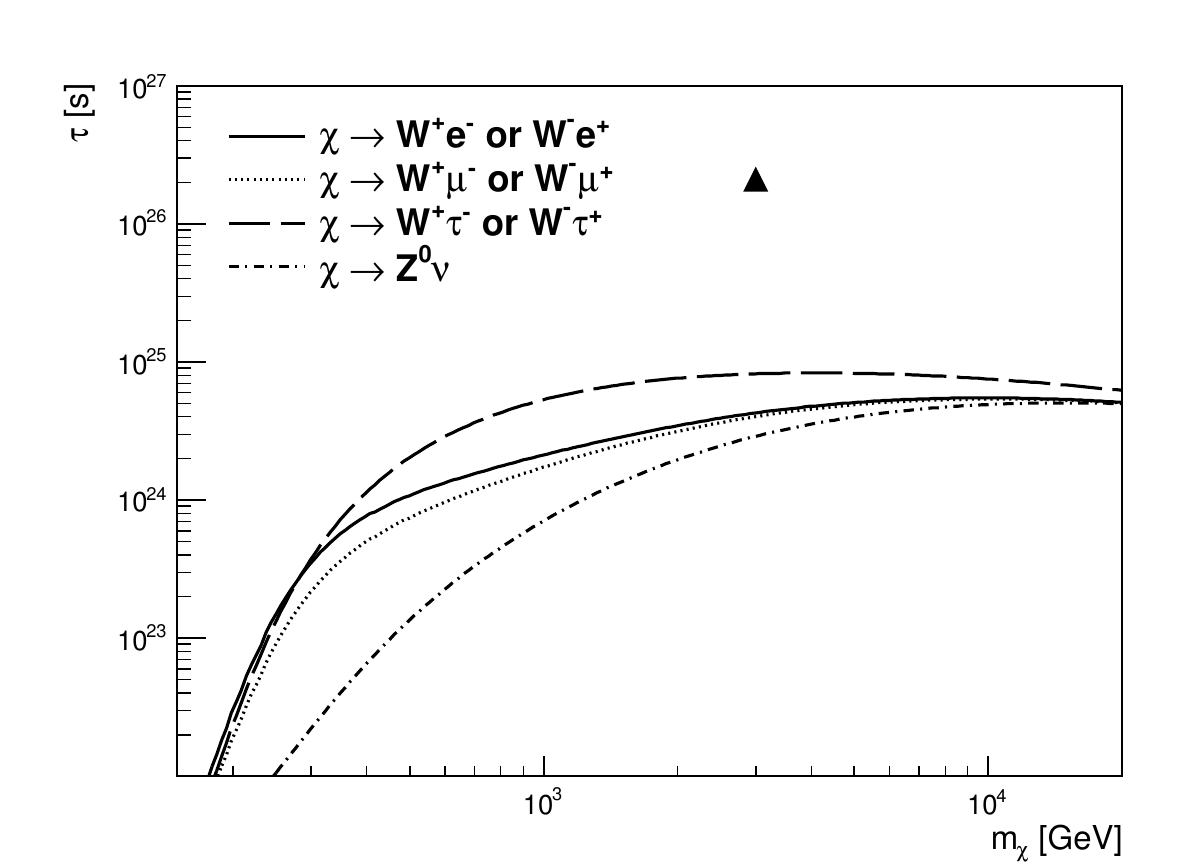}
             }
   \caption{95\% CL LLs from the VERITAS observations of Segue 1 on the decay lifetime as a function of the dark matter particle mass. Left: Bosonic dark matter decaying to two identical particles: $\mathrm{W^{+}W^{-}}$, $\mathrm{b\bar{b}}$, $\mathrm{\tau^{+}\tau^{-}}$, $\mathrm{e^{+}e^{-}}$ and $\mathrm{\mu^{+}\mu^{-}}$. The black star and the black cross denote the best fits to the Fermi and PAMELA data considering the $\mathrm{\mu^{+}\mu^{-}}$ and $\mathrm{\tau^{+}\tau^{-}}$ channels, respectively, and are taken from \cite{DDMPam5}.  Right: Fermionic dark matter decaying to two different particles: $\mathrm{W^{\pm}\ell^{\mp}}$ (where $\mathrm{\ell = e,\,\mu,\,\tau}$) and $\mathrm{Z^{0}\nu}$. The black triangle indicates the best fit to the Fermi and PAMELA data considering the channel $\mathrm{W^{\pm}\mu^{\mp}}$, taken from  \cite{DDMPam5}.}
   \label{Fig4}
 \end{figure*}

If we assume that dark matter is a decaying particle, LLs on the lifetime of dark matter can be derived. In decaying dark matter scenarios, the dark matter particle can either be bosonic or fermionic. The LLs are computed using eq. \ref{Ngamma} and making the appropriate substitutions to eq. \ref{dPhidE_DM}, as explained in section \ref{DMModel}. For bosonic dark matter particles, the same channels as in the annihilating dark matter case are considered: $\mathrm{W^{+}W^{-}}$, $\mathrm{b\bar{b}}$, $\mathrm{\tau^{+}\tau^{-}}$, $\mathrm{e^{+}e^{-}}$ and $\mathrm{\mu^{+}\mu^{-}}$. The decay spectra are the same as those used for the annihilating dark matter bounds (see right panel of Figure \ref{Fig2}, and eq. \ref{LeptonSpectra}), making the substitution for the scaled variable $\mathrm{x \rightarrow 2x}$, or equivalently $\mathrm{m_{\chi} \rightarrow m_{\chi}/2}$. The left panel of Figure \ref{Fig4} shows the 95\% LLs on the decay lifetime $\mathrm{\tau}$ for the five channels mentioned above. The limits peak at the level of $\mathrm{\tau \sim 10^{24}-10^{25}\,s}$, depending on the dark matter particle mass.\\
In the case of fermionic dark matter, the dark matter particle decays to different channels. The following channels are considered for the exclusion limits: $\mathrm{W^{\pm}\ell^{\mp}}$ (where $\mathrm{\ell = e,\,\mu,\,\tau}$) and $\mathrm{Z^{0}\nu}$. Each corresponding $\gamma$-ray spectrum is a combination of the same spectra used in section \ref{sigmav} for the computation of limits on annihilating dark matter. They have been simulated with the PYTHIA 8.1 package \cite{PYTHIA}. The right side of Figure \ref{Fig4} shows the corresponding LLs on $\mathrm{\tau}$, which peak in the range $\mathrm{\tau \sim 10^{24}-10^{25}\,s}$.\\
Interestingly, decaying dark matter models have been suggested to be good alternatives to annihilating dark matter models for explaining the ATIC \cite{ATIC} and PAMELA \cite{PAMELApositronfraction} lepton anomalies, because they circumvent model-building issues such as the {\it ad hoc} addition of boost factors (e.g Sommerfeld enhancement, and/or astrophysical boost factors, see section \ref{leptons}). Spectral fits to the Fermi-LAT and PAMELA data prefer channels involving hard lepton spectra such as $\mathrm{\mu^{+}\mu^{-}}$, $\mathrm{\tau^{+}\tau^{-}}$ and $\mathrm{W^{\pm}\mu^{\mp}}$, with a dark matter particle in the mass range 2-5 TeV and with a lifetime of the order of $\mathrm{\tau \sim 10^{26}\,s}$ \cite{DDMPam1,DDMPam4,DDMPam5}. The VERITAS limits on the dark matter particle decay lifetime are at least an order of magnitude away from the best fits to the Fermi and PAMELA data (see Figure \ref{Fig4}) and do not rule out these models.
\section{Testing the dark matter interpretation of the cosmic ray lepton anomalies}\label{leptons}
The excess in the cosmic ray electron spectrum measured by the ATIC collaboration \cite{ATIC}, and the unexpected rise of the positron fraction \cite{PAMELApositronfraction} in conjunction with the lack of anti-proton excess \cite{PAMELAantiproton} reported by the PAMELA collaboration, have received considerable attention over the past few years. Even if these features can easily be explained by conventional astrophysical sources \cite{ElectronAstroSource1,ElectronAstroSource2}, the dark matter interpretation has been extensively studied and has led to numerous dark matter models. In any dark matter annihilation interpretation, the cosmic ray lepton excesses require a dark matter particle mostly annihilating into leptons and with a significant $\mathrm{{\cal{O}}(10^{2}-10^{3})}$ boost to the thermal freeze-out annihilation cross-section $\mathrm{\langle\sigma v\rangle \sim 3\times10^{-26}\,cm^{3}\,s^{-1}}$. The substructures residing in the Milky Way dark matter halo do not provide the necessary boost factor \cite{Clumps3} and are unlikely to be responsible for the lepton anomalies \cite{ElectronAstroBoost}. However, models including a Sommerfeld enhancement to the annihilation cross-section naturally solve this issue. In this section, we use the VERITAS Segue 1 observations to test such models and, more generally, to derive limits on the boost factor in a model-independent way.
\subsection{Models with a Sommerfeld enhancement}
\begin{figure*}[!ht]
   \centerline{\includegraphics[width=3.3in]{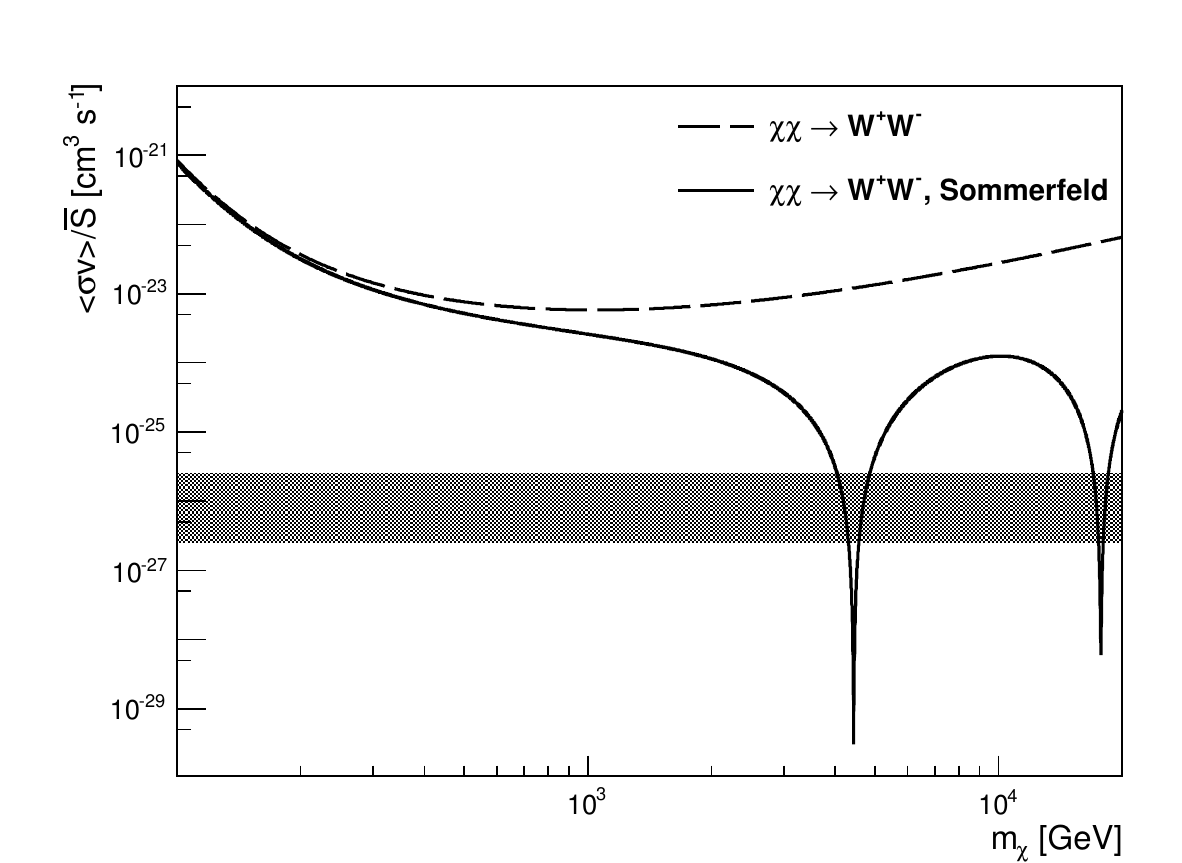}
              \hfil
              \includegraphics[width=3.3in]{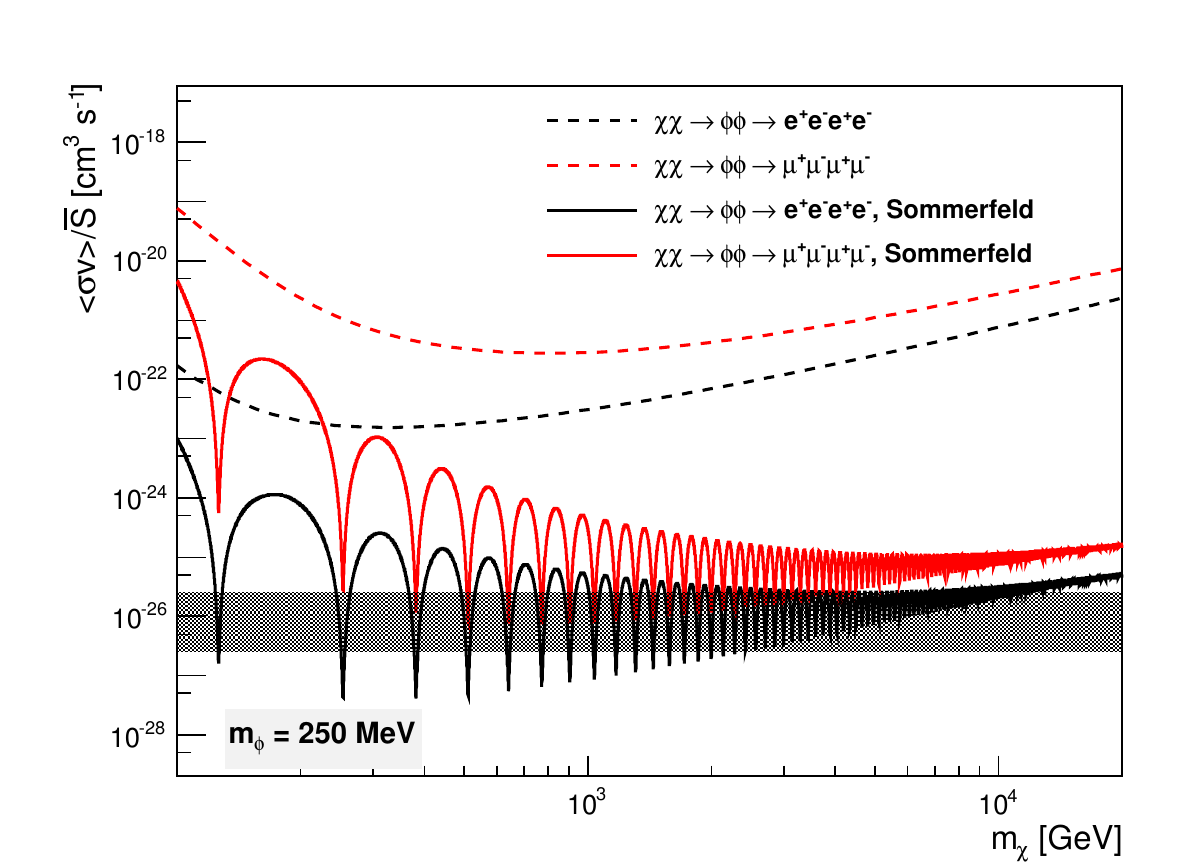}
             }
   \caption{95\% CL exclusion curves from the VERITAS observations of Segue 1 on $\mathrm{\langle \sigma v \rangle}$/$\mathrm{\bar{S}}$ as a function of the dark matter particle mass, in the framework of two models with a Sommerfeld enhancement. The expected Sommerfeld enhancement $\mathrm{\bar{S}}$ applied to the particular case of Segue 1 has been computed assuming a Maxwellian dark matter relative velocity distribution. The grey band area represents a range of generic values for the annihilation cross-section in the case of thermally produced dark matter. Left: model I with wino-like neutralino dark matter annihilating to a pair of $\mathrm{W^{+}W^{-}}$ bosons. Right: model II with a 250 MeV scalar particle decaying into either $\mathrm{e^{+}e^{-}}$ or $\mathrm{\mu^{+}\mu^{-}}$. See text for further details.}
   \label{Fig5}
 \end{figure*}

The Sommerfeld enhancement is a non-relativistic quantum effect arising when two dark matter particles interact through an attractive potential \cite{Sommerfeld3}, mediated by a particle $\mathrm{\phi}$. The Sommerfeld correction S (or Sommerfeld boost) is velocity-dependent ($\mathrm{S\sim1/v}$) and modifies the product of the annihilation cross-section and the relative velocity:
\begin{equation}
\mathrm{\sigma v = S(v,m_{\chi},m_{\phi},\alpha) \times (\sigma v)_{0},}\label{eqSom}
\end{equation}
where $\mathrm{(\sigma v)_{0}}$ is the WIMP annihilation cross-section times its relative velocity at thermal freeze-out. As shown by eq. \ref{eqSom}, the Sommerfeld correction also depends on the dark matter particle mass $\mathrm{m_{\chi}}$, the mass $\mathrm{m_{\phi}}$ of the particle mediating the attractive potential, and its coupling $\alpha$ to the dark matter particle. Depending on the mass and the coupling of the exchanged particle, the Sommerfeld enhancement can exhibit a serie of resonances for specific values of the dark matter particle mass, giving very large boost factors up to $\mathrm{10^{6}}$ \cite{Sommerfeld3}. The Sommerfeld enhancement is of particular interest for cold dark matter halos like dSphs, where the mean dark matter velocity dispersion can be as low as a few $\mathrm{km\,s^{-1}}$. The computation of the Sommerfeld enhancement for a relative velocity $\mathrm{v}$, a dark matter particle mass $\mathrm{m_{\chi}}$, a mediator mass $\mathrm{m_{\phi}}$ and a coupling $\mathrm{\alpha}$ is usually performed through the numerical resolution of the Schrodinger equation, modeling the attractive potential with a Yukawa potential \cite{Sommerfeld3}. Instead of numerically solving the Schr\"{o}dinger equation for each set of parameters ($\mathrm{v}$,$\mathrm{m_{\chi}}$,$\mathrm{m_{\phi}}$,$\mathrm{\alpha}$), we use an analytic solution by approximating the Yukawa potential with the Hulth\'{e}n potential \cite{Hulthen1,Hulthen2}. The analytic solution has been shown to closely match the numerical solution \cite{Hulthen2}. In order to calculate the Sommerfeld enhancement in Segue 1, one has to average the Sommerfeld boost factor over the dark matter relative velocity distribution $\mathrm{f(v)}$:
\begin{equation}
\mathrm{\bar{S}(m_{\chi},m_{\phi},\alpha) = \int S(v,m_{\chi},m_{\phi},\alpha) f(v)\,dv.}
\end{equation}
The velocity-weighted annihilation cross-section in the dark matter halo is then given by \cite{Hulthen2}:
\begin{equation}
\mathrm{\langle \sigma v \rangle = (\sigma v)_0 \times \bar{S}(m_{\chi},m_{\phi},\alpha).}
\end{equation}
Following \cite{Segue1HL1}, the Segue 1 dark matter relative velocity distribution is assumed to be Maxwellian, i.e. the dark matter gas is thermalized and at equilibrium, with a mean relative velocity dispersion of $\mathrm{v_{0} \simeq 6.4\,km\,s^{-1}}$.\\
In this section, we focus on two models comprising a Sommerfeld enhancement to the annihilation cross-section. The first model (hereafter model I, \cite{Sommerfeld3}) assumes that the dark matter particle is a wino-like neutralino $\mathrm{\chi^{0}}$, arising in SUSY extensions of the standard model. To circumvent the helicity suppression of the annihilation cross-section into light leptons, the neutralino can oscillate with charginos $\mathrm{\chi^{\pm}}$, which themselves can preferentially annihilate into leptons. The transition to a chargino state is mediated by the exchange of a $\mathrm{Z^{0}}$ boson ($\mathrm{m_{Z^{0}} \sim 90\,GeV}$, $\mathrm{\alpha \sim 1/30}$), leading to a Sommerfeld enhancement. The second model (hereafter model II) introduces a new force in the dark sector \cite{AH}. The new force is carried by a light scalar field $\mathrm{\phi}$ predominantly decaying into leptons and with a mass $\mathrm{{\cal{O}}(1\,GeV)}$ and coupling to standard model particles chosen to prevent the overproduction of antiprotons. In such models, dark matter annihilates to a pair of $\mathrm{\phi}$ scalar particles, with an annihilation cross-section boosted by the Sommerfeld enhancement. The coupling $\mathrm{\alpha}$ of the light scalar particle $\mathrm{\phi}$ to the dark matter particle is determined assuming that $\mathrm{\chi\chi \rightarrow \phi\phi}$ is the only channel that regulates the dark matter density before freeze-out \cite{Hulthen2}.\\
Figure \ref{Fig5} shows the VERITAS constraints for each of these models, derived with the observations of Segue 1. The dashed curves show the 95\% CL exclusion limits without the Sommerfeld correction to the annihilation cross-section, whereas the solid curves are the limits to the Sommerfeld enhanced annihilation cross-section. The left panel of Figure \ref{Fig5} shows the constraints on model I, for the annihilation of neutralinos into $\mathrm{W^{+}W^{-}}$ through the exchange of a $\mathrm{Z^{0}}$ boson. The Sommerfeld enhancement exhibits two resonances in the considered dark matter particle mass range, for $\mathrm{m_{\chi} \simeq 4.5\,TeV}$ and $\mathrm{m_{\chi} \simeq 17\,TeV}$, respectively. VERITAS excludes these resonances, which boost the annihilation cross-section far beyond the canonical $\mathrm{\langle \sigma v \rangle \sim 3\times10^{-26}\,cm^{3}\,s^{-1}}$. The right panel of Figure \ref{Fig5} shows the VERITAS constraints on model II, for a scalar particle with mass $\mathrm{m_{\phi} = 250\,MeV}$. The Sommerfeld enhancement exhibits many more resonances, located at different dark matter particle masses and with different amplitudes with respect to model I, because the coupling and mass of the exchanged particle differ. Two channels in which the scalar particle decays either to $\mathrm{e^{+}e^{-}}$ or $\mathrm{\mu^{+}\mu^{-}}$ have been considered. VERITAS observations start to disfavor such models, especially for the $\mathrm{e^{+}e^{-}e^{+}e^{-}}$ channel where some of the resonances are beyond $\mathrm{\langle \sigma v \rangle \sim 3\times10^{-26}\,cm^{3}\,s^{-1}}$. This result holds for $\mathrm{\phi}$ particle masses up to a few GeV.
\\
\subsection{Model-independent constraints on the boost factor}
\begin{figure}[!h]
    \centering
  \includegraphics[width=3.5in]{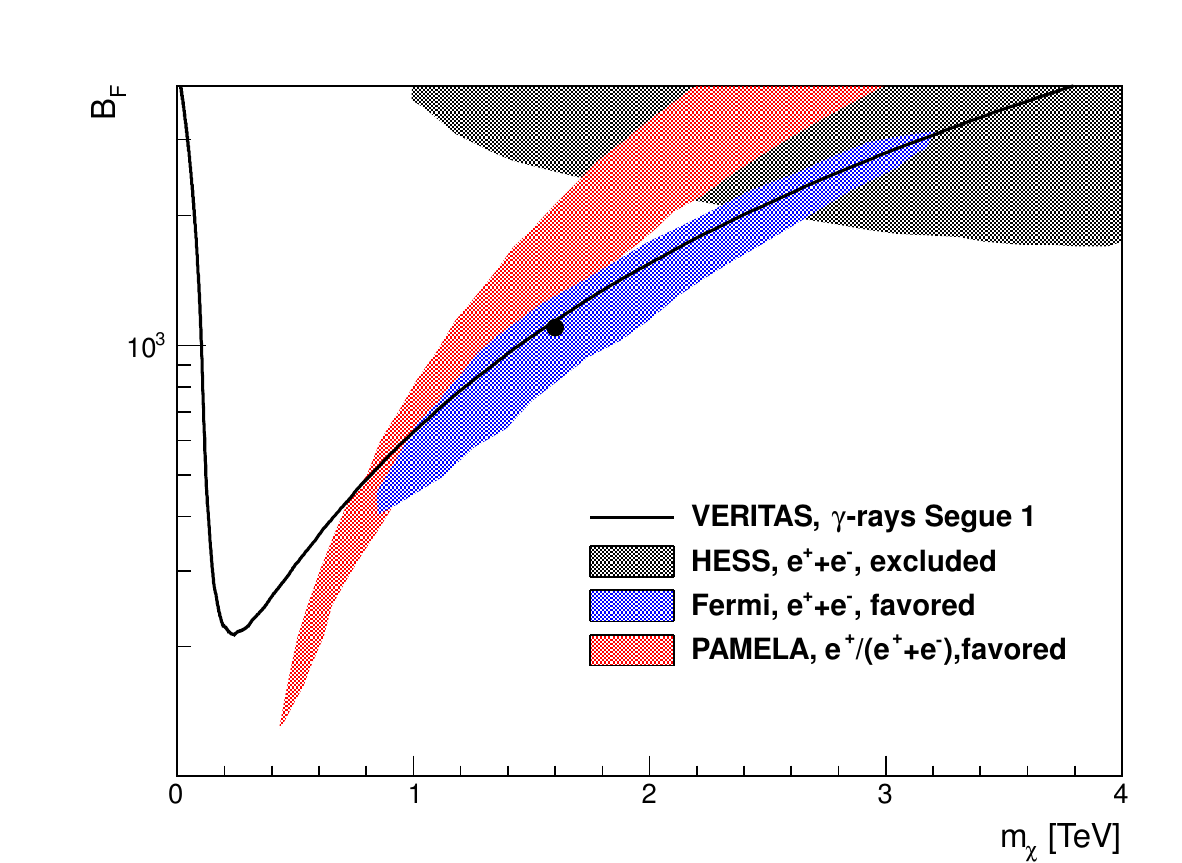}
  \caption{95\% CL exclusion limits on the overall boost factor $\mathrm{B_{F}}$ as a function of the dark matter particle mass from the VERITAS observations of Segue 1, assuming that the dark matter particles annihilate exclusively into $\mathrm{\mu^{+}\mu^{-}}$. The shaded areas are the $\mathrm{2\,\sigma}$ contours derived from fits to the H.E.S.S., PAMELA and Fermi-Lat data. The black dot is an example of a model that simultaneously fits well the H.E.S.S., PAMELA, and Fermi-LAT data. See \cite{BergstromElectron} for further details.}
  \label{Fig6}
 \end{figure}

In the previous section, we have explicitly constrained the Sommerfeld boost factor to the annihilation cross-section in the framework of two interesting models. Here, an example of model-independent constraints on the overall boost factor $\mathrm{B_{F}}$ (particle physics and/or astrophysical boost) as a function of the dark matter particle mass is presented. The constraints are then compared to the recent cosmic ray lepton data.\\
Following \cite{BergstromElectron}, we assume that dark matter annihilates exclusively into muons with an annihilation cross-section $\mathrm{\langle\sigma v \rangle = 3\times 10^{-26}\,cm^{3}\,s^{-1}}$. In such a case, we use the dashed exclusion curve of Figure \ref{Fig3} (right) to compute 95\% limits on $\mathrm{B_{F}}$. Figure \ref{Fig6} shows the 95\% CL ULs on the overall boost factor $\mathrm{B_{F}}$. The blue and red shaded regions are the 95\% CL contours that best fit the Fermi-LAT and PAMELA $\mathrm{e^{+}e^{-}}$ data, respectively. The grey shaded area shows the 95 \% CL excluded region derived from the H.E.S.S. $\mathrm{e^{+}e^{-}}$ data \cite{BergstromElectron}. The black dot is an example of a model which simultaneously fits well the H.E.S.S., PAMELA and Fermi-LAT data. The VERITAS VHE $\gamma$-ray observations of Segue 1 rule out a significant portion of the regions preferred by cosmic ray lepton data. However, the electron and positron constraints depend on the cosmic ray propagation model, especially on the electron energy loss parameter $\mathrm{\tau_0}$ \cite{BergstromElectron}. The Klein-Nishina suppression of the inverse Compton loss rate can significantly alter this parameter at energies above 100 GeV \cite{Epropag}. In such a case, the Fermi and PAMELA 95\% CL contours would scale down, and the VERITAS limits would then be relaxed.
\\
\section{Conclusions}\label{Conclusion}
The ground-based VHE $\mathrm{\gamma}$-ray observatory VERITAS has started an extensive observation campaign toward the nearby Segue 1 dSph, one of the most dark matter-dominated satellite galaxies currently known. With nearly 48 hours of exposure, we derived strong flux ULs and constraints on the annihilation cross-section or decay life time of a hypothetical WIMP, independently of any dark matter models. The reported integral flux ULs at the 95\% CL are at the level of 0.5\% of the Crab Nebula flux above a minimum energy of 300 GeV. The corresponding limits on the velocity-weighted annihilation cross-section are in the range $\mathrm{\langle\sigma v \rangle^{95\%\,CL} \leq 2-9\times 10^{-24}\,cm^{3}\,s^{-1}}$, depending on the annihilation channel considered. These are the most constraining limits reported so far with any dSph observations conducted in the VHE $\gamma$-ray band. The limits are complementary to those provided by the Fermi-LAT collaboration \cite{dSphFermi}, but are at least two orders of magnitude away from the canonical value $\mathrm{\langle \sigma v \rangle \sim 3\times10^{-26}\,cm^{3}\,s^{-1}}$. Bounds on the lifetime of decaying dark matter have also been derived, with LLs in the range $\mathrm{\tau^{95\%\,CL} \geq 10^{24}-10^{25}\,s}$, an order of magnitude below the models that best fit the electron and positron excesses recently measured in cosmic ray spectra.  Finally, the VERITAS Segue 1 data have also been used to test the dark matter interpretation of the cosmic ray lepton anomalies. The Segue 1 data disfavors annihilating dark matter models with a Sommerfeld enhancement, confirming the results of \cite{dSph4} and \cite{Hulthen2}. Furthermore, the VERITAS observations start to exclude scenarios where the dark matter annihilates preferentially into a $\mathrm{\mu^{+}\mu^{-}}$ pair, which is the favored scenario for good fits to the H.E.S.S., PAMELA and Fermi-LAT electron and positron data \cite{BergstromElectron}.\\
The uncertainties on the dark matter limits derived throughout this paper mostly come from the modeling of the Segue 1 dark matter density profile. As opposed to the classical dSphs, the lack of a high statistics star sample for Segue 1 prevents an accurate modeling of the dark matter distribution. Assuming an Einasto profile, the systematic uncertainties in the dark matter profile modeling can change the astrophysical factor, and hence they scale up or down the limits by a factor of 4 at the $\mathrm{1\,\sigma}$ level \cite{Segue1HL1}. The systematic uncertainties could even be larger if the Segue 1 dark matter profile is compatible with a cored profile. Future spectroscopic surveys might increase the Segue 1 star sample and reduce the uncertainties on its dark matter content.\\
\section*{Acknowledgments}
We thank Rouven Essig, Neelima Sehgal and Louis E. Strigari for useful discussions about the Segue 1 dark matter distribution profile. VERITAS is supported by grants from the US Department of Energy Office of Science, the US National Science Foundation, and the Smithsonian Institution, by NSERC in Canada, by Science Foundation Ireland (SFI 10/RFP/AST2748), and by STFC in the UK. We acknowledge the excellent work of the technical support staff at the Fred Lawrence Whipple Observatory and at the collaborating institutions in the construction and operation of the instrument.

\newpage
\includepdf[pages={1}]{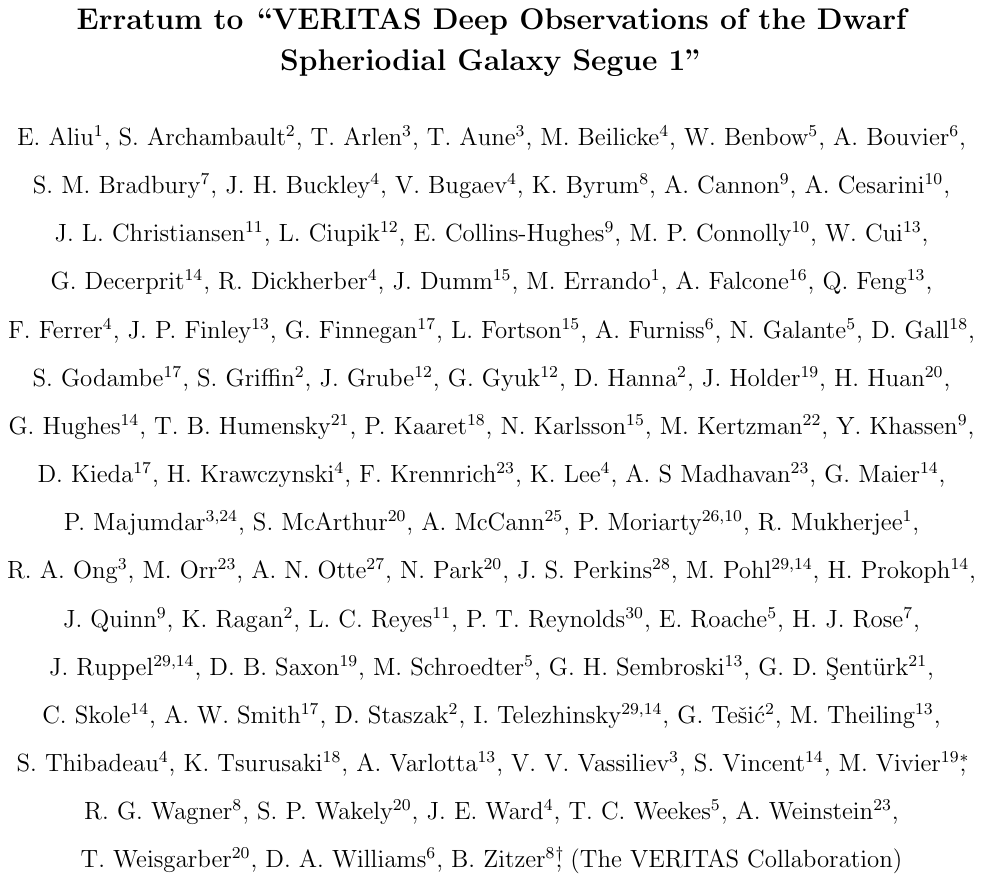}
\includepdf[pages={2}]{erratum_segue1_v4.pdf}
\includepdf[pages={3}]{erratum_segue1_v4.pdf}
\includepdf[pages={4}]{erratum_segue1_v4.pdf}
\includepdf[pages={5}]{erratum_segue1_v4.pdf}
\includepdf[pages={6}]{erratum_segue1_v4.pdf}
\includepdf[pages={7}]{erratum_segue1_v4.pdf}
\includepdf[pages={8}]{erratum_segue1_v4.pdf}

\end{document}